 \documentclass[smallabstract,smallcaptions]{dccpaper}

\usepackage{epsfig}
\usepackage{citesort}
\usepackage{amsmath}
\usepackage{amssymb}
\usepackage{color}
\usepackage{url}

\usepackage{multirow}
\usepackage{booktabs}
\usepackage{cite}
\newlength{\figurewidth}
\newlength{\smallfigurewidth}

\begin{document}

\title
{\large
\textbf{Point Cloud-Assisted Neural Image Compression}
}

\author{%
Ziqun Li$^{1}$, Qi Zhang$^{1}$, Xiaofeng Huang$^2$, Zhao Wang$^2$, Siwei Ma$^2$, and Wei Yan$^{1, \ast}$\\[0.5em]
{\small\begin{minipage}{\linewidth}\begin{center}
\begin{tabular}{cc}
$^{1}$Peking University  & $^{2}$Advanced Institute of Information Technology, Peking University\\
\end{tabular}
\end{center}\end{minipage}}
}

\maketitle
\thispagestyle{empty}

\begin{abstract}
High-efficient image compression is a critical requirement. In several scenarios where multiple modalities of data are captured by different sensors, the auxiliary information from other modalities are not fully leveraged by existing image-only codecs, leading to suboptimal compression efficiency. In this paper, we increase image compression performance with the assistance of point cloud, which is widely adopted in the area of autonomous driving. We first unify the data representation for both modalities to facilitate data processing. Then, we propose the point cloud-assisted neural image codec (PCA-NIC) to enhance the preservation of image texture and structure by utilizing the high-dimensional point cloud information. We further introduce a multi-modal feature fusion transform module (MMFFT) to capture more representative image features, remove redundant information between channels and modalities that are not relevant to the image content. Our work is the first to improve image compression performance using point cloud and achieves state-of-the-art performance.
\end{abstract}

\Section{Introduction}

As the volume of image data escalates exponentially, the constraint of limited bandwidth necessitates the development of highly efficient compression methodologies. Across diverse industries, particularly those embracing multimodal data, such as autonomous driving, virtual reality, and smart city applications, the demand for advanced image compression techniques is arising. These domains typically integrate both image and point cloud. However, current image compression methods predominantly focus solely on optimizing the image compression performance by using images as input, neglecting the potential benefits of utilizing all available modal data. These single modality-only approaches cannot achieve the optimal performance, so we want to use point cloud to assist in image compression.

Image compression techniques have been developed for decades. Traditional standards such as JPEG \cite{Pen92}, BPG \cite{Bellard15}, WebP \cite{Google10}, and VVC \cite{Bross21} offer appreciable compression efficiency. 
However, they are limited by hand-craft design in each module, lacking overall optimization capabilities. 
Unlike traditional methods, learned-based approaches undertake end-to-end optimization. The pioneering end-to-end image codec, conceived by Ballé et al. \cite{Ballé16}, laid the foundation for subsequent advancements. This groundbreaking work employed a Convolutional Neural Network (CNN)-based architecture, intricately structured with key modules encompassing analysis transform, quantization, entropy model, synthesis transform, and entropy coding. Subsequent research has built upon this robust framework, refining and enhancing its capabilities. Ballé et al. \cite{Ballé18} introduced a notable improvement in the form of a variational autoencoder-based hyper-prior model, which ingeniously incorporated side information and harnessed the power of the univariate Gaussian distribution for hyper-prior modeling. Some works \cite{Minnen18,Cui21,Cheng20,Liu20} further refined this approach by adopting different Gaussian distribution for the hyper-prior. The entropy model facilitates the estimation of the probability distribution of compressed image features, thereby enabling efficient encoding of these features. 
Some works \cite{Ballé16,Guo21} proposed context coding to capture correlations among symbols. To address the computational inefficiency of these context models, He et al. \cite{He21} devised a checkerboard context model, and Minnen et al. \cite{Minnen20} introduced channel context model and segmented the latent representation $\widehat{y}$ into slices. Subsequently, various enhanced entropy models \cite{He22,Jiang23,Li24,Sui24} were successively proposed, incrementally enhancing the efficiency of image compression. 
 In pursuit of enhancing image compression efficiency beyond merely refining the entropy model, several studies have ventured into the utilization of diverse neural network architectures. He et al. \cite{He16} proposed a residual network to solve the problems of gradient vanishing and exploding in deep network training and improve performance. Some works \cite{Zhu22,Zou22,Qian21,Koyuncu22} attempted to construct the learned image codec based on the Transformer. Liu et al. \cite{Liu23} efficiently incorporate the local modeling ability of CNN and the non-local modeling ability of Transformer to enhance the ability to capture image features.
Attention modules help the model focus on important regions. Some works \cite{Liu19,Cheng20,Zou22,Li24,Sui24} used their proposed attention modules to improve image compression performance. 

However, in fields that contain multiple modal data, we naturally want to use the correlation between different modalities to assist in image compression. The above learning-based models typically extract features from the two-dimensional textural and contextual information of the image, employing successive downsampling to mitigate redundancy. Despite leveraging the two-dimensional properties of the image, these models neglect potential insights derived from three-dimensional viewpoints. Compared to images, point cloud offers a high-dimensional perspective, furnishing supplementary spatial structural information that is not readily discernible from the image alone, thereby facilitating more effective compression. Integrating point cloud data allows the neural network to reinterpret images from a more holistic perspective, enhancing the efficiency of feature extraction and achieving superior compression performance. In light of this, we propose a novel point cloud-assisted neural image compression in this paper. We first unify the representation of image and point cloud for data processing purposes. Then, we design PCA-NIC to use point cloud as auxiliary information to assist in image compression. Furthermore, we design MMFFT to empower the neural network with the ability to learn richer and more discriminative image features. This module serves three purposes. The first is that it sharpens the model's focus on common features shared between image and point cloud within individual channels while pruning redundant information. The second is that it encourages the model to capture more comprehensive image features. The last is that it meticulously filters out features that are irrelevant to the image, further refining the feature representation. Our contributions are summarized as follows:

\begin{itemize}
\item{We propose the first point cloud-assisted neural image codec, PCA-NIC. By incorporating point cloud data as auxiliary information, this model enhances the retention of image details and the expression of structural information, significantly improving compression efficiency and image quality.} 
\item{We propose a unified digital representation for image and point cloud data, mapping the point cloud data into the pixel coordinate system of the image and jointly representing them as 4×H×W array. The unified representation facilitates information fusion and reduces the complexity of data conversion and processing.} 
\item{We propose a novel multi-modal feature fusion transform module, MMFFTM, which concatenates the features of image and point cloud along the channel dimension. It employs a channel attention mechanism to increase the weight of similar parts while reducing the impact of the point cloud distinct from the image regions.}
\item{Extensive experiments demonstrate that our approach achieves state-of-the-art performance on KITTI dataset. PCA-NIC outperforms cheng2020 \cite{Ballé16} by 54.518\% in Bjøntegaard-delta-rate (BD-rate).} 
\end{itemize}

\Section{Proposed Method}
\label{headings}

\SubSection{Analysis of Enhanced Image Compression Performance with Point Cloud Assistance}

\begin{figure}[ht]
\centering
\includegraphics[width=3in]{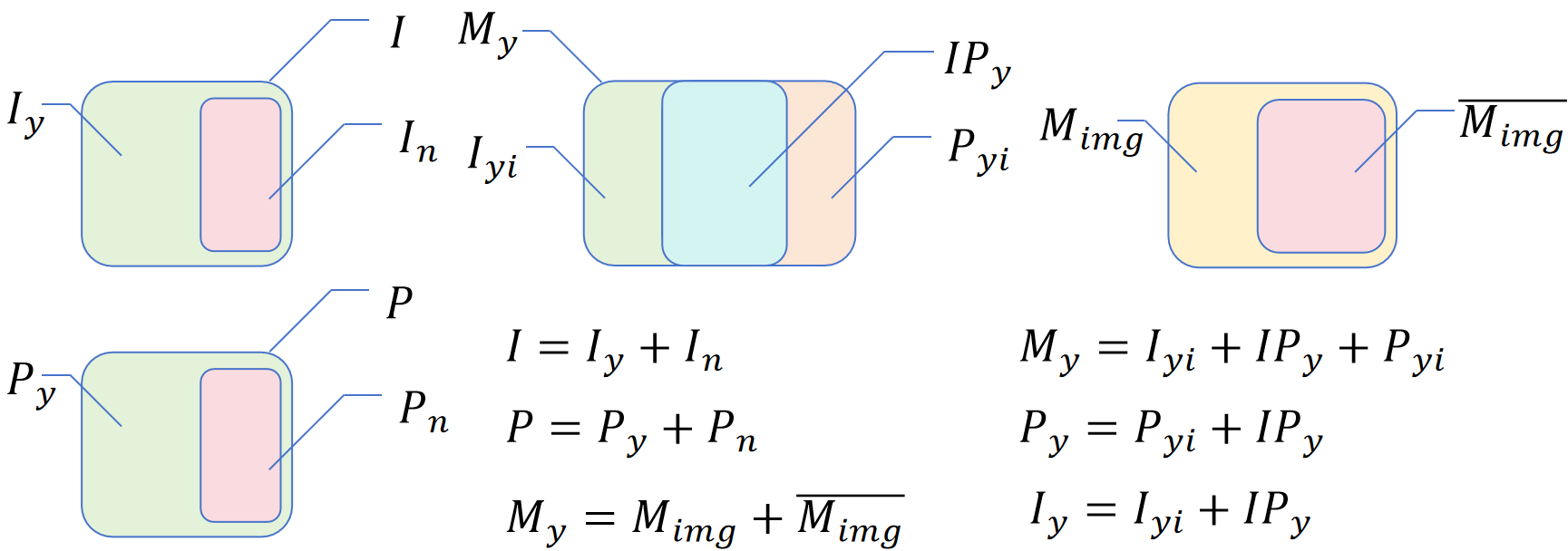}
\caption{Assuming all features of the image are $I$, the image features that can be extracted by the neural network are $I_y$, and the image features that cannot be extracted by the neural network are $I_n$. Similarly, regarding point cloud features, we have $P$, $P_y $, and $P_n$. By using neural networks to simultaneously extract features from both image and point cloud, the total mixed feature that can be extracted is $M_y$, the shared feature between the extracted image and point cloud is $IP_y$, the feature that only belongs to the image is $I_{yi}$, and the feature that only belongs to the point cloud is $P_{yi}$. The features belonging to the image in the mixed feature $M_y$ extracted by the neural network and the point cloud are denoted as $M_{img}$, while the features not belonging to the image are denoted as $\overline{M_{img}}$.}
\label{entropy}
\end{figure}

\begin{figure}[ht]
\centering
\includegraphics[width=4in]{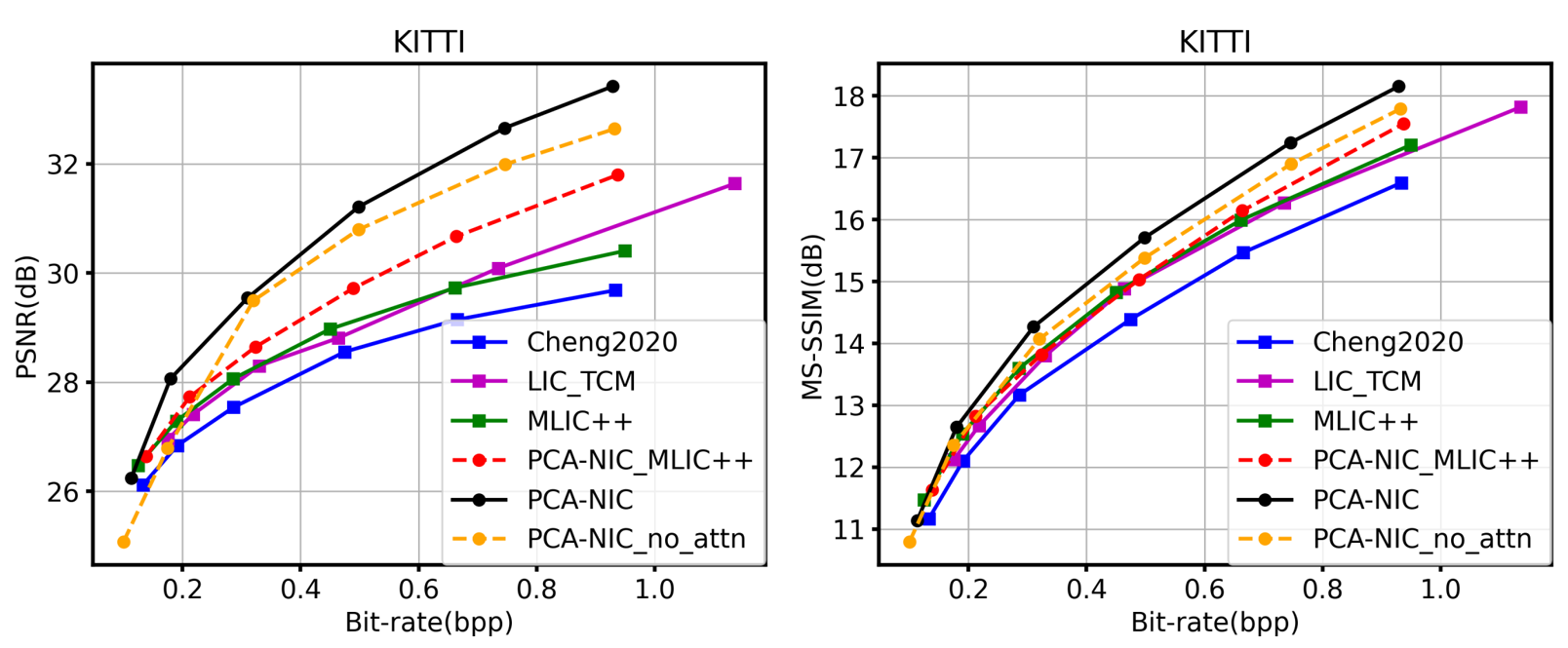}
\caption{PSNR-Bit-rate curve and MS-SSIM-Bit-rate curve.}
\label{psnr}
\end{figure}

We first analyze the potential benefits of point cloud-assisted image compression from the perspective of information theory. In Fig. \ref{entropy}, we show the symbols of features extracted from image and point cloud as well as their correlations. Then, the information entropy of the entire image features, features can be extracted from the image, and features cannot be extracted are denoted as $H(I)$, $H(I_y)$, and $H(I_n)$, respectively. Obviously,  
\begin{equation}
\label{deqn_ex1a}
H(I) = H(I_y) + H(I_n).
\end{equation}
The information entropy of the features extracted from point cloud is $H(P_y)$. The overlapping features between the extractable image features and point cloud features have an information entropy of  $H(IP_y)$, with the distinct information entropies being $H(I_{yi})$ and $H(P_{yi})$. Then, the information entropy of the mixed features extracted from both image and point cloud can be calculated as 
\begin{equation}
\label{deqn_ex1a}
H(M_y) = H(I_y,P_y) = H(I_y) + H(P_y|I_y).
\end{equation}
Since $H(IP_y) > 0$, we have $H(P_y|I_y) > 0$ and $H(M_y) > H(I_y)$. This indicates that compared to extracting image features solely from images, extracting image features from mixed data of point cloud and images enables neural networks to understand image features from a larger range or more comprehensive perspective. Obviously, the latter will be easier to understand features that the former cannot or is difficult to comprehend. Our experimental results (refer to Fig. \ref{psnr} and Table \ref{BD-Rate}) also confirm this viewpoint. From another perspective, the entropy value of features extracted from images in mixed data is greater than the entropy value of features extracted from images alone, that is $H(M_{img})>H(I_y)$. 
This indicates that an increasement in the diversity of input data will enhance the ability of CNN to extract specified features.

\SubSection{Unified Representation of Image and Point Cloud} 
\label{aaa}
Typically, image can be presented by a C×H×W array, where C,H,W are the number of channels, image width, and image height, respectively. On the other hand, point cloud is represented as an N×M array, where N is the number of points and M encompasses features like spatial positions (X, Y, Z), reflections, or color attributes (R, G, B). Due to the unordered nature of points in N × M point cloud, it is difficult to obtain the relative positional relationships between points in space through neural networks. Moreover, the different representation forms of these two modalities can bring difficulties to data processing. Therefore, it is necessary to unify their representation.

\begin{figure}[ht]
\centering
\includegraphics[width=4in]{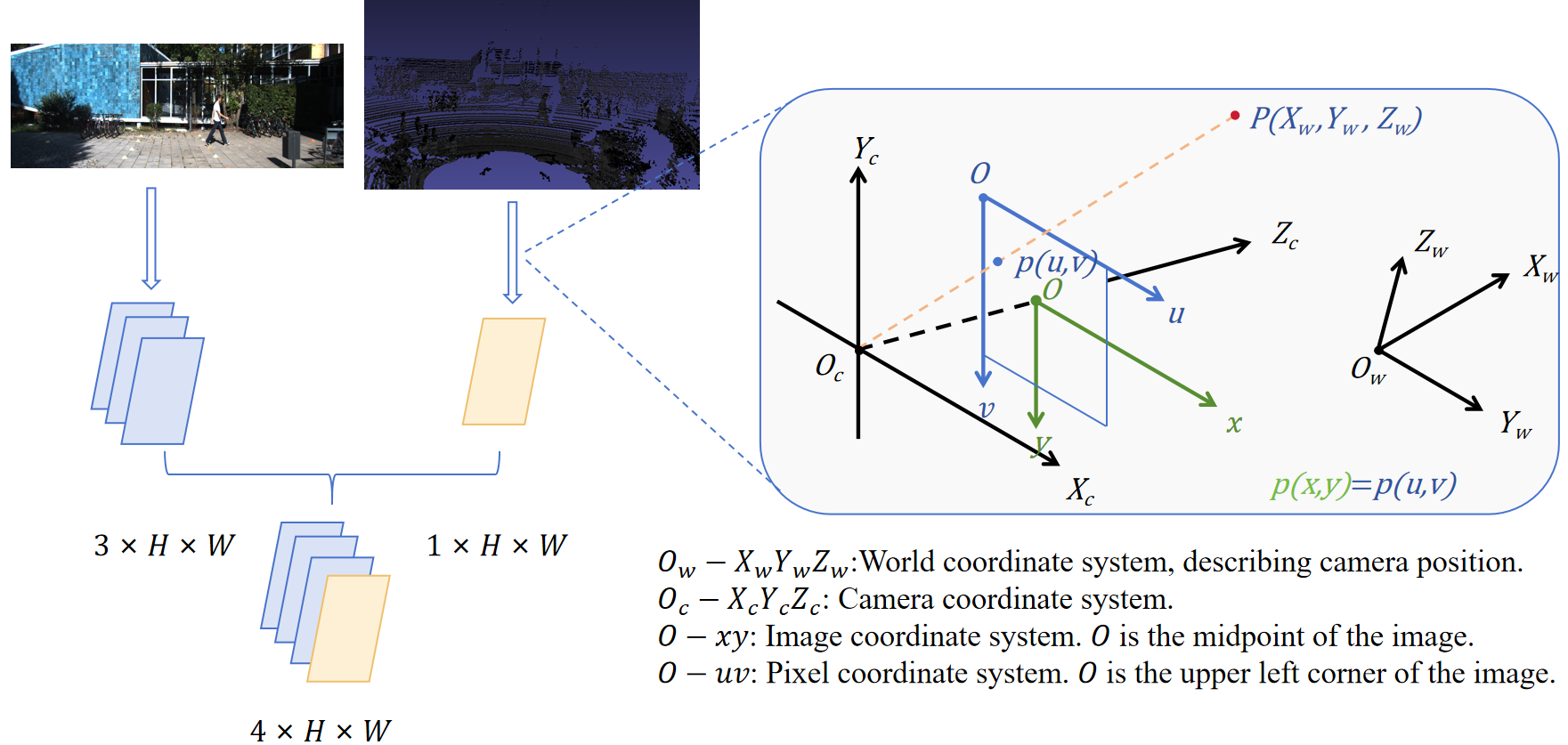}
\caption{The unified representation of image and point cloud. The  projection is to transform $P(X_w,Y_w,Z_w)$ into $p(u,v)$.}
\label{projecting}
\end{figure}

\begin{equation}
\label{formula_projecting}
X_{2D} = P_{rect}^{(2)} R_{rect}^{(0)} Tr_{velo}^{cam} X_{3D}.
\end{equation}

As shown in Fig. \ref{projecting}, we project point cloud onto an image to transform unordered points into ordered ones, and explicitly represent the relative positional relationships between points. We use the KITTI dataset as an example to explain how to perform the aforementioned operations. 
The KITTI dataset contains image, point cloud and a calibration file. The calibration file includes the rotation transform matrix $T_{velo}^{cam}$ from radar to camera, the rotation correction matrix $R_{rect}^{(0)}$ for the camera, the corrected camera projection matrix $P_{rect}^{(2)}$. As shown in Formula (\ref{formula_projecting}) \cite{Geiger13}, $X_{3D}$ denotes the input point cloud, which is a 3×N array. The final output is a 3×N array $X_{2D}$, obtained by dividing the third row by the first two rows to yield the final $X_{2D}$. The first row of data, u, corresponds to the image width W, while the second row, v, corresponds to the image height H. This successfully converts the three-dimensional point cloud data into a two-dimensional pixel coordinate system. 

Most importantly, the depth information of the point cloud (represented by the first row of the 3×N matrix $X_{3D}$) is used as the value of the points in the pixel coordinate system. This results in a point cloud data in the form of 1×H×W. By concatenating the image and point cloud data along the channel dimension, a unified representation of the image and point cloud is obtained, which is 4×H×W.

\SubSection{PCA-NIC}

\begin{figure}[ht]
\centering
\includegraphics[width=5.5in]{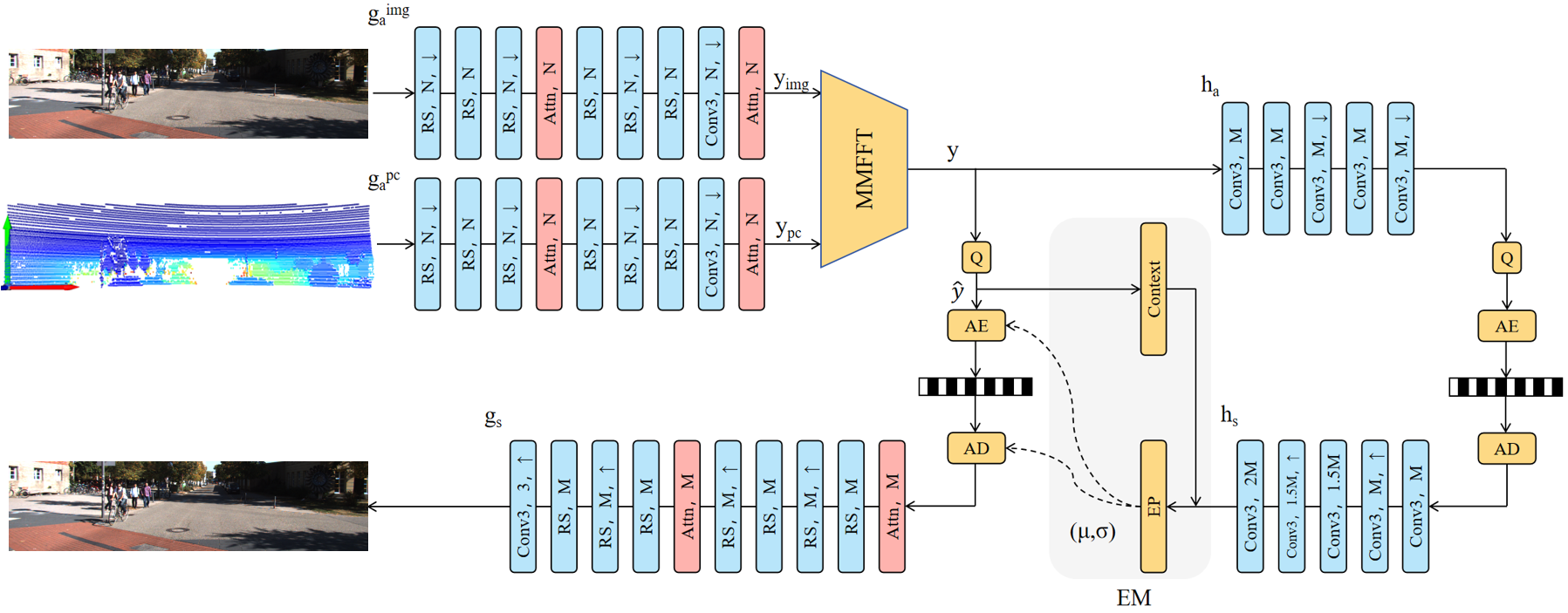}
\caption{The overall architecture of PCA-NIC. $\downarrow$ means down-sampling. $\uparrow$ means up-sampling. RS is residual network. Attn is the attention module of \cite{Cheng20}. N and M are channels, where N and M is equal to 192 and 288, respectively.}
\label{frame}
\end{figure}

As depicted in Fig. \ref{frame}, we first give an overview of PCA-NIC. Given that point cloud data and images have been converted into a unified digital representation similar to images, i.e. 4 × H × W, we can use image feature extraction networks to extract point cloud features. We design two analysis transform modules to extract image and point cloud features: the image analysis transform module $g_{a}^{img}$ and the point cloud analysis transform module $g_{a}^{pc}$. We propose MMFFT to use point cloud to supplement image features. To provide additional edge information, we use a hyper prior analysis module $h_a$ and a hyper prior synthesis module $h_s$. The entire architecture also includes a quantization module (Q), an entropy model (EM), and a synthesis transformation module $g_s$. $g_a^{img}$, $g_s$, $g_a^{pc}$, $h_a$ and $h_s$ enhance the module of \cite{Cheng20} by modifying channel counts for optimized point cloud feature extraction, boosting model capacity, and facilitating improved semantic processing of multimodal data. MMFFT takes the image latent representation $y^{img}$ obtained through $g_a^{img}$ and the point cloud latent representation $y^{pc}$ obtained through $g_a^{pc}$ as inputs, and then outputs a fused image and point cloud latent representation $y$. Context coding employs the methodology detailed in work of \cite{Cheng20}. Entropy estimation is based on a Gaussian mean scale distribution.

\SubSection{MMFFT}

\begin{figure}[ht]
\centering
\includegraphics[width=2.5in]{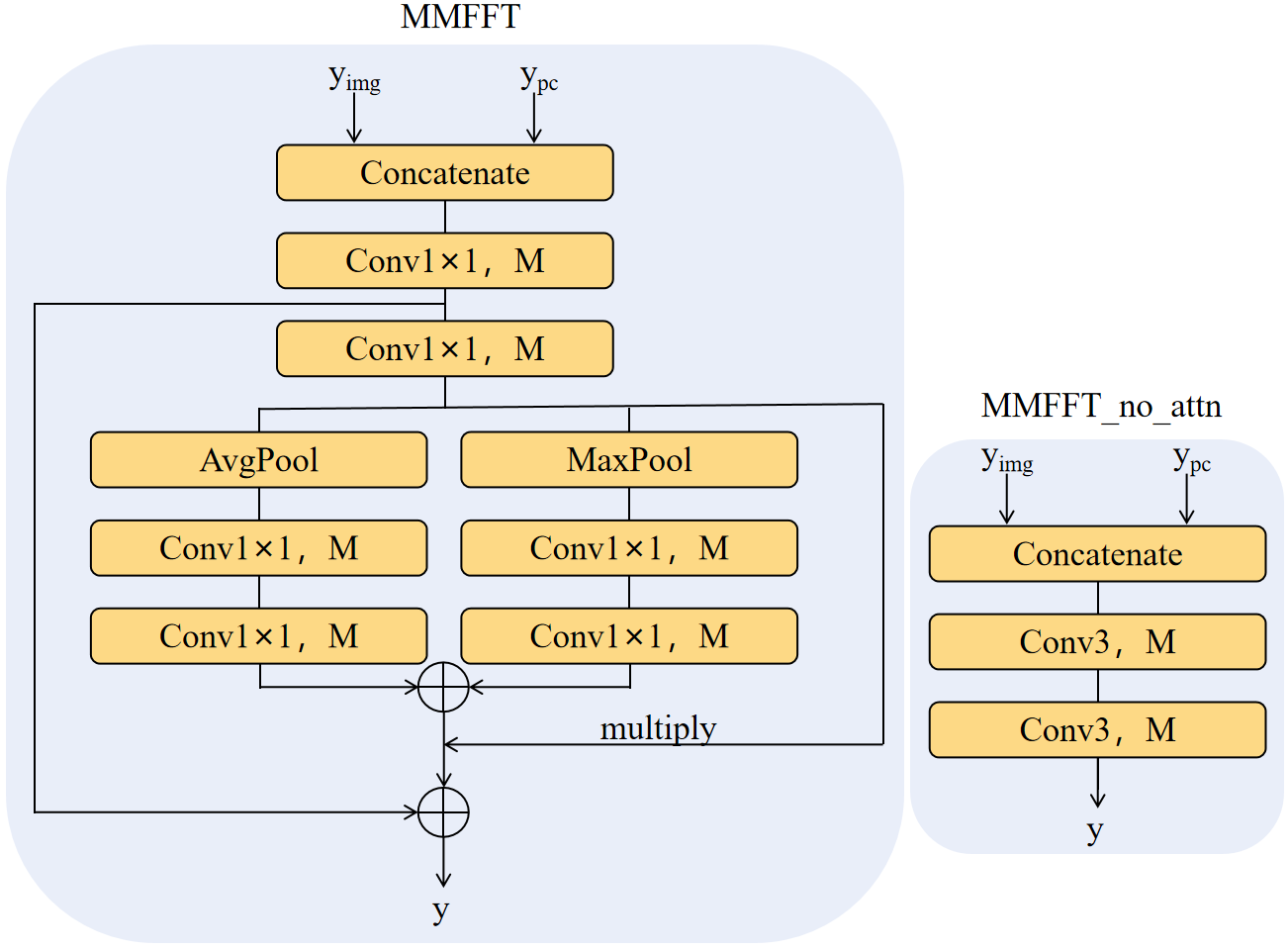}
\caption{The left part is the diagram of MMFFT, and the right is the frature fusion transform without attention mechanisms (MMFFT\_no\_attn).}
\label{FFTM}
\end{figure}

When describing the same scene from the same viewpoint, both image and point cloud data share common semantic information, such as the scene's structure and object shapes. They also exhibit differences, notably in lighting conditions. To enhance feature extraction, we propose a novel module that optimizes the learning process of neural networks by increasing the weight of similar components and minimizing the impact of dissimilarities. 
As depicted in Fig. \ref{FFTM}, the inputs of MMFFT are $y_{img}$ and $y_{pc}$. They are concatenated along the channel dimension to form a 2N×H×W array. This array then changes the number of channels from 2N to M through a convolution. After that, the data stream is divided into two paths. One is used to construct residuals, and the other is used to construct channel attention modules through average pooling and maximum pooling. Add these two processed data together to obtain $y$. 

MMFFT identifies the overlapping features between image and point cloud, assigning higher weights to these shared features. Consequently, the neural network focuses more on the common semantic information during the training phase. After assigning higher attention to shared features, only one shared feature will be retained and duplicate features will be removed, which can further reduce the bitrate.
$y$ contains impurities unrelated to the image, such as noise and image independent features from point cloud. To mitigate the negative effects of these impurities, we only compare the reconstructed image with the original image when optimizing. This encourages the neural network to reduce impurities during the learning process, thereby enhancing the accuracy of feature extraction. 
The image semantic information contained in y obtained through MMFFT is more abundant compared to the image semantic information obtained through feature extraction using only images. This is because by introducing point cloud data, the diversity of feature extraction by neural networks is increased, enabling them to understand and extract image features from a more comprehensive perspective.

\Section{Experiments}
\label{others}
\SubSection{Training Settings}

We used the KITTI dataset for training and testing, which consisted of 7,481 training samples and 7,518 testing samples. We cropped the image to a size of 256×256 pixels, and we cropped the point cloud corresponding to the image. 
The testing dataset comprised randomly cropped 256×256 pixel samples from the first 5000 examples in the KITTI dataset's testing samples. For training models with different levels of distortion, we used NVIDIA GeForce RTX 4090. We employed MSE as distortion metrics, with $\lambda$ values for MSE set to ${0.0016, 0.0032, 0.0075, 0.015, 0.03, 0.045}$. We aimed for the largest possible batch size, with the learning rate starting from 1e-4 and decreasing to 1e-8 until the test loss plateaued, with the number of epochs typically ranging between 100 and 150.

\SubSection{Results}

\begin{table}[tp]
\begin{center}
\caption{\label{BD-Rate}%
BD-PSNR and BD-rate.}
{
\renewcommand{\baselinestretch}{1}\footnotesize
\begin{tabular}{lcccc}
\toprule 
\multicolumn{1}{l}{~}&
\multicolumn{2}{c}{\textbf{PSNR}} &
\multicolumn{2}{c}{\textbf{MS-SSIM}}\\

\multicolumn{1}{l}{\textbf{Method}} &
\textbf{BD-PSNR (dB)} & \textbf{BD-rate (\%)} & \textbf{BD-SSIM} & \textbf{BD-rate (\%)}\\
\midrule
Cheng2020          & 0          & 0           & 0          & 0           \\  
LIC\_TCM           & 0.416      & -19.594     & 0.397      & -13.142     \\  
MLIC++             & 0.525      & -23.63      & 0.504      & -14.194     \\  
PCA-NIC\_MLIC++    & 0.991      & -34.784     & 0.467      & -15.474     \\  
PCA-NIC\_no\_attn   & 1.73       & -39.905     & 0.708      & -21.719     \\  
PCA-NIC           & 2.101      & -54.518     & 1.001      & -29.315     \\  
\bottomrule
\end{tabular}}
\end{center}
\end{table}

\begin{figure}[ht]
\centering
\includegraphics[width=5in]{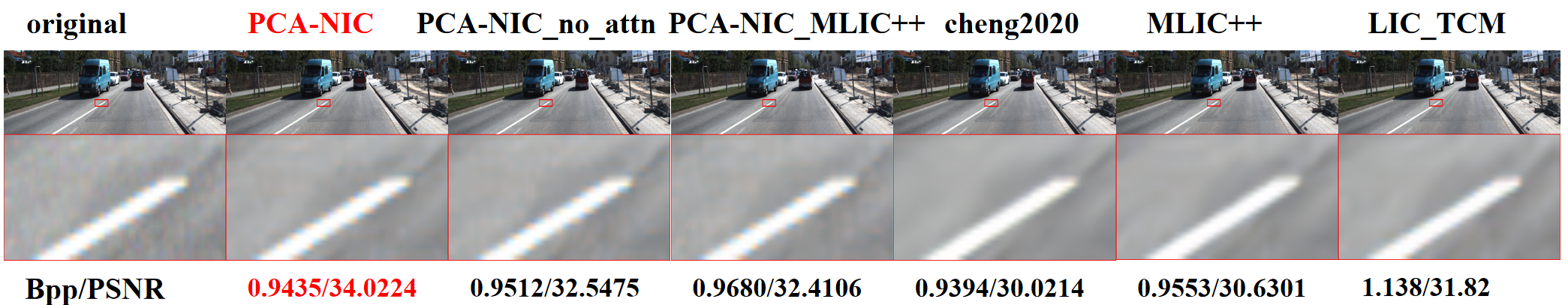}
\caption{Visualization of the reconstructed 007517.png from the KITTI dataset.}
\label{visualization}
\end{figure}

The validation set is comprised of randomly cropped 256×1024-sized data from the regions where the height is 110 pixels or above, within the last 2518 samples of the test data. Fig. \ref{psnr} shows the rate-distortion performance on KITTI. We report the BD-PSNR, BD-SSIM and BD-rate in Table \ref{BD-Rate}. Fig. \ref{visualization} shows a comparison of the effects of using different codecs to decode an image in KITTI. Our proposed method (PCA-NIC) achieves state-of-the-art performance on KITTI when measured in PSNR and MS-SSIM. PCA-NIC reduces BD-rate by 54.518\% on KITTI over Cheng2020 \cite{Cheng20} when measured in PSNR and a 29.315\% decrease in MS-SSIM. PCA-NIC achieves an improvement of 2.101dB in BD-PSNR on KITTI over Cheng2020 \cite{Cheng20} when measured and a 1.001 increase in BD-SSIM. According to the comparison results, it can be concluded that the improvement in compression performance between the two is attributed to PCA-NIC. We also proposed two models, one (PCA-NIC\_no\_attn) without attention mechanism in feature fusion transform module to investigate the effectiveness of our proposed MMFFT module (refer to next subsection), and the other (PCA-NIC\_no\_attn) with the entropy model of MLIC++ that includes global, local, and channel attention mechanisms to explore the role of our entropy model module (refer to next subsection). The results show that, PCA-NIC\_no\_attn and PCA-NIC\_no\_attn have better compression performance than \cite{Cheng20}, but worse than PCA-NIC. This demonstrates the effectiveness of our MMFFT and our entropy model in improving compression performance.

\SubSection{Ablation Studies}

\textit{Settings}. We conducted corresponding ablation studies and evaluated the contributions of MMFFT and our entopy model on the selected KITTI validation set. The learning rate was set to $10^{-4}$, and the training was terminated when the loss function value showed negligible change. The training and testing datasets during the training process were consistent with those mentioned previously. The results are presented in Fig. \ref{psnr} and Table \ref{BD-Rate}. 

\textit{Analysis of MMFFT}. To investigate the role of the MMFFT module, we conducted a comparative experiment between the MMFFT module with attention mechanism and the MMFFT\_no\_attn module (refer to Fig. \ref{FFTM}) without attention mechanism. The results show that the MMFFT module can help neural networks understand and capture image features from a more comprehensive perspective. The MMFFT module can make neural networks pay more attention to the common features of image and point cloud, and only retain one common feature. MMFFT can eliminate features or impurities unrelated to images. The MMFFT module greatly improves image compression performance. 

\textit{Analysis of entropy model}. To verify that our entropy model is more suitable for image compression utilizing point cloud, we conducted a comparative experiment using the entropy model of MLIC++ \cite{Jiang23} instead of our entropy model. The results (refer to Fig. \ref{psnr} and Table \ref{BD-Rate}) show that our entropy model is indeed more suitable for point cloud assisted image compression. Furthermore, by comparing PCA-NIC\_MLIC++ with MLIC++, we can see the superiority of our entropy model of PCA-NIC.

\Section{Conclusion}
\label{others}

We have proposed a novel image codec PCA-NIC that leverages point cloud spatial information to enrich image texture and structure. PCA-NIC is the first to improve image compression performance using point cloud and achieves state-of-the-art performance. The meaning of our work is as follows: Firstly, it pioneers point cloud application in image compression, opening up a new direction for image processing and compression. Secondly, it promises wide use in VR, autonomous driving, smart cities, and drone detection, enhancing data processing efficiency and accuracy. Thirdly, it enriches image compression theory, providing fresh insights for future research. Challenges include high computational complexity, point cloud data accuracy impacting compression, and opportunities for further algorithm optimization. Future research should focus on developing more efficient algorithms that can speed up processing times, enhance the accuracy of point cloud data acquisition and processing, and improve compression performance by leveraging depth information, as well as by implementing improved fusion and entropy modules.

\Section{References}
\bibliographystyle{IEEEbib}
\bibliography{main}

\end{document}